\documentclass{emulateapj}
\usepackage{apjfonts}
\usepackage{natbib}
\usepackage{amssymb, amsmath, amsbsy, epsfig, epsf}%, slashbox, rotating}
\usepackage[dvips]{color}

\def\obj{RX\,J1633+4718}
\def\rosat{{\it ROSAT}}
\def\iras{{\it IRAS}}
\def\iso{{\it ISO}}
\def\chisq{$\chi^2$}

\def\ulum{${\rm erg\,s^{-1}}$}
\def\mbh{$M$}
\def\mbhx{$M_{\rm x}$}
\def\msun{\ensuremath{M_{\odot}}}
\def\mdot{$\dot{m}$}
\def\nh{$N_{\rm H}$}
\def\gnh{$N_{\rm H}^{\rm Gal}$}
\def\unh{${\rm cm^{-2}}$}
\def\kmps{${\rm km\,s^{-1}}$ }

\def\redd{$L/L_{\rm Edd}$}
\def\hb{H$\beta$}
\def\ha{H$\alpha$}

\def\tbb{$T_{\rm bb}$}
\def\rsch{$R_{\rm s}$}
\def\tmax{$T_{\rm max}$}
\def\ktmax{$kT_{\rm max}$}
\def\rin{$R_{\rm in}$}
\def\rini{$R_{\rm in}\sqrt{\cos \theta}$}

\def\tinc{$T'_{\rm in}$}
\def\ktinc{$kT'_{\rm in}$}
\def\rinc{$R'_{\rm in}$}
\def\rinci{$R'_{\rm in}\sqrt{\cos \theta}$}

\slugcomment{} \shorttitle{Signature of a black hole accretion disc}
\shortauthors{Yuan W. et al.}

\begin{document}

\title{X-ray observational signature of a black hole  accretion disc
in an active galactic nucleus RX\,J1633+4718}

\author{
W. Yuan\altaffilmark{1,2}, .
B.F. Liu\altaffilmark{1,2},
H. Zhou\altaffilmark{3},
T.G. Wang\altaffilmark{3}
}

\altaffiltext{1}{National Astronomical Observatories/Yunnan Observatory,
Chinese Academy of Sciences, Kunming, Yunnan, P.O. BOX 110, China}

\altaffiltext{2}{Key Laboratory for the Structure and Evolution of Celestial
Objects, Chinese Academy of Sciences}

\altaffiltext{3}{Key Laboratory for Research in Galaxies and Cosmology,
Center for Astrophysics, University of Science and Technology of China,
Hefei, Anhui, China}

\email{wmy@ynao.ac.cn}

\begin{abstract}

We report the discovery of a luminous ultra-soft X-ray excess
in a radio-loud narrow-line Seyfert\,1 galaxy, RX\,J1633+4718,
from archival \rosat\ observations.
The thermal temperature of this emission, when fitted with a blackbody,
is as low as $32.5^{+8.0}_{-6.0}$\,eV.
This is in remarkable contrast to the
canonical temperatures of $\sim$0.1--0.2\,keV
found hitherto for the soft X-ray excess in active galactic nuclei (AGN),
and is interestingly close to the  maximum temperature
predicted for a postulated accretion disc in this object.
If this emission is indeed blackbody in nature,
the derived luminosity ($3.5^{+3.3}_{-1.5}\times10^{44}$\,\ulum)
infers a compact emitting area with a size
($\sim5\times10^{12}$\,cm or 0.33\,AU in radius)
that is comparable to several times the Schwarzschild radius
of a black hole at the mass
 estimated for this AGN ($\sim3\times10^6$\,\msun).
In fact, this ultra-steep X-ray emission can be well fitted as the
(Compton scattered)  Wien tail of the multi-temperature blackbody emission
from an optically thick accretion disc,
whose parameters inferred
(black hole mass and accretion rate)  are in good agreement with
independent estimates using optical emission line spectrum.
We thus consider this feature as a signature of the long-sought
X-ray radiation directly from a disc around a super-massive black hole,
presenting observational evidence for
a black hole accretion disc in AGN.
Future observations with better data quality, together with improved
independent measurements of the black hole mass, may constrain the spin of
the black hole.
\end{abstract}


\section{Introduction}

Accretion of matter onto a black hole (BH) via an
optically thick  disc \citep{ss73,nt73}
is widely accepted  as the engine to power active galactic nuclei (AGN) and
black hole X-ray binaries (BHXBs) in a bright state.
In observation, a search of evidence for
BH accretion discs has long been pursued,
and one important signature to look for is
the  predicted characteristic multi-temperature blackbody radiation.
For BHXBs (BH masses \mbh$\sim10$\,\msun)
the bulk of the disc blackbody emission,
which has a maximum temperature \ktmax$\sim$0.5--1\,keV,
falls within the readily observable X-ray bandpass,
and the disc model has proven to be prevailing
\citep[see e.g.][for a recent review]{done07a}.
In AGNs habouring super-massive black holes (SMBH) with typical
\mbh$\sim10^{6-9}$\,\msun\ and hence much lower \tmax, however,
the situation is much less clear.
The main difficulty lies in that the bulk of the
disc emission falls within the far-to-extreme UV regime, which
is not observable due to Galactic absorption.
%% the next sentence can be removed if needed.
While it is generally thought that the optical/UV "big blue bump" of AGN
is likely associated with disc emission \citep[e.g.][]{shi78,mal83},
some key issues have not been settled yet, however \citep[e.g.][]{laor97}.

A long-standing puzzle is that,
although the (Compton scattered) Wien tail of the disc blackbody emission
is expected to plunge into the soft X-ray band \citep[e.g.][]{ross92,st95a},
convincing evidence remains illusive.
Indeed, a soft X-ray excess is commonly detected in AGNs;
however, their thermal temperatures, which appear to fall within a narrow range
($kT\sim$0.1--0.2\,keV),
are too high to conform the disc model prediction \citep[e.g.][]{gd04},
even with the effects of Compton scattering taken into account \citep{ross92}.
Neither the inferred luminosities agree with the model values self-consistently
\citep[e.g.][]{min09}.
Furthermore, the independence of temperature on \mbh\ found recently
 argues against the disc blackbody origin \citep[e.g.][]{bian09}.
In general, the soft X-ray excess is suggested to
have different origins from disc emission \citep[e.g.][]{done07b}.
Although there were limited previous attempts to link the soft X-ray excess
with disc emission in a few AGNs,
e.g.\ Mkn766 \citep[][]{mol93}
and RE\,J1034+396 \citep[][]{puc01},
whereby \mbh\ and accretion rates were constrained from the X-ray
(and optical/UV) data,
the evidence is far from convincing.

In this paper, we report on an ultra-soft
X-ray excess detected in an AGN \obj,
whose thermal temperature ($kT\sim$30\,eV) is among the lowest ever detected.
We show that this emission is well consistent with blackbody radiation from
an optically thick accretion disc around a SMBH.
We assume a cosmology with $H_{0}$= 70 km\, s$^{-1}$\,Mpc$^{-1}$,
$\Omega_{M}=0.3$, and $\Omega_{\Lambda}=0.7$.
Errors are quoted at the 90\% confidence level unless mentioned otherwise.

\section{RX\,J1633+4718}

\obj\ (SDSS\,J163323.58+471859.0)
was first identified with a narrow-line Seyfert\,1 (NLS1) galaxy
at a redshift $z$=0.116 in optical identification of the
\rosat\ All-Sky Survey (RASS) sources \citep{moran96,wis97}.
The AGN resides in one of a  pair of galaxies, with the other
a starburst galaxy  separated by 4\arcsec\ \citep{bade95}.
It is associated with a variable, invert spectrum radio source \citep{neum94}
with an unresolved VLBI core at milli-arcsec resolution \citep{doi07}.
It is among a very radio-loud NLS1 AGN sample studied by \citet{yuan08}.
This rare type of AGNs exhibit properties characteristic of blazars,
i.e.\ having relativistic jets directed
 close to the line-of-sight \citep{zhou07,yuan08},
 which has been confirmed in several of the objects by recent $\gamma$-ray
observations with {\it Fermi}  \citep{abdo09}.

Its optical spectrum, acquired in the Sloan Digital Sky Survey (SDSS) and
analysed in \citet[][see their Figure\,1]{yuan08},
has broad components of the \hb\ and \ha\ lines
with a width $FWHM=909\pm43$\,\kmps\ and
a luminosity 3.14(11.7)$\times10^{41}$\,\ulum, respectively.
The expected "thermal" continuum luminosity at $5100\AA$
{\em estimated from the \hb\ luminosity}
is $\lambda L_{\lambda5100}=2.63\times10^{43}$\,\ulum\
%($2.73\times10^{43}$\,\ulum\ from \ha).
using the relations from \citet{gh05}.
The "thermal" bolometric luminosity
 (i.e.\ free from the non-thermal jet emission)
can then be estimated as
$L_{\rm bol}$=$k\lambda L_{\lambda5100}$
=$2.6(\pm 0.5)\times 10^{44}$\,\ulum, adopting
$k=10\pm2$ ($1\sigma$) for {\em radio-quiet} AGNs \citep{rich06}.
The BH mass, estimated from the broad Balmer linewidth and
the luminosity of either
the broad line or the "thermal" continuum
lies within \mbh$\simeq 2.0-3.5\times10^6$\,\msun\
using various commonly adopted formalisms \citep[e.g.][]{vp06,col06,gh07}.
We  thus adopt a nominal value \mbh$=3\times10^6$\,\msun,
with a systematic uncertainty of $\sim0.3$\,dex \citep{vp06}.
The estimated Eddington ratio is then \redd=$0.69^{+0.73}_{-0.35}$,
using the above "thermal" bolometric luminosity.

\section{Ultra-soft X-ray excess emission}

J1633+4718 was observed with the
ROSAT PSPC-b on axis with a  3732\,s exposure on July 24th, 1993,
and also detected during the RASS (PSPC-c) with a weighted exposure of  909\,s.
The reduction and a brief analysis of the \rosat\ data
was presented in \citet{yuan08}.
To summarise, there are $\sim976$ net source counts
 in the pointed observation and 185 counts in the RASS.
The 0.1--2.4\,keV spectrum can be well modeled with two components,
an ultra-soft X-ray component dominating energies  below 0.4\,keV and
a flat power-law (photon index $\Gamma\sim1.37$) dominating energies above 0.4\,keV;
the latter was interpreted as
inverse Compton emission from a relativistic jet \citep{yuan08}.
In this paper we focus on the ultra-soft X-ray component and
revisit the spectrum by performing a more detailed analysis.

\begin{figure}
%\epsscale{0.8}
\includegraphics[width=0.5\hsize,angle=0]{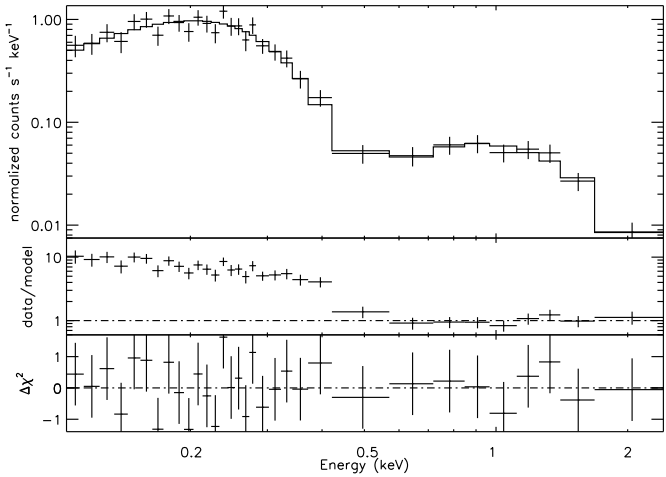}
\caption{
Upper panel: \rosat\ PSPC spectrum of \obj\ and the best-fit model
of a
power-law plus a blackbody (\nh\ fixed at Galactic \gnh).
Middle: data to model ratio, where a power-law model
with fixed \gnh\  is fitted to the 0.5--2.4\,keV band and extrapolated down to 0.1\,keV.
An overwhelming soft X-ray excess is evident.
Lower: residuals of the best-fit model as in the upper panel.
}
\label{fig:xspec}
\end{figure}

We use the pointed observation spectrum (Figure\,\ref{fig:xspec})
to derive spectral parameters,
given its much higher data signal-to-noise ratio (S/N) than that of the RASS data.
{\it XSPEC} (v.12.5) is used to perform spectral fit.
Neutral absorption with a \ion{H}{1} column density \nh\
is always added in  spectral models concerned.
The Galactic \gnh=$1.79\times10^{20}$\,\unh\ on the line-of-sight,
 measured from the LAB Survey \citep{Kal05}.
The results are summarised in Table\,1.
We first try a simple power-law model.
The fit is unacceptable for fixing \nh=\gnh\ (\chisq/d.o.f.=67/30),
underestimating severely the fluxes at both the low and high energies.
Setting \nh\ free still yields a poor fit,
since the same systematic structures remain in the  residuals and,
even worse, resulting in little or no absorption.
Fitting the higher-energy (0.4--2.4\,keV) spectrum only
(fixing \nh= \gnh) yields a good fit and a flat $\Gamma=1.63$($\pm0.4$),
while extrapolating this power-law to lower energies
reveals a prominent soft X-ray excess, as shown in
Figure\,\ref{fig:xspec} (middle panel).

Therefore an additional emission component is added to
account for the soft X-ray excess, that
is  modeled by commonly adopted models
including blackbody, power-law, thin plasma and thermal bremstrahlung.
In general, all these models yield nearly the same excellent fits
and the systematic structures in the residuals vanish.
However, they cannot be distinguished based on fitting statistics.
For most of the models, and particularly blackbody,
the fitted \nh\ values are in good agreement
with \gnh, albeit large uncertainties.
We thus fix the absorption \nh=\gnh\
when deriving spectral parameters (Table\,1).
As an example,
the best-fit power-law plus blackbody model
and the residuals are shown in Figure\,\ref{fig:xspec}.
We note that the addition of a second, relatively steep
power-law continuum as in normal Seyferts with
fixed $\Gamma=$2--3 is not required,
indicating that it is negligible.

\begin{figure}
\includegraphics[width=0.4\hsize,angle=-90]{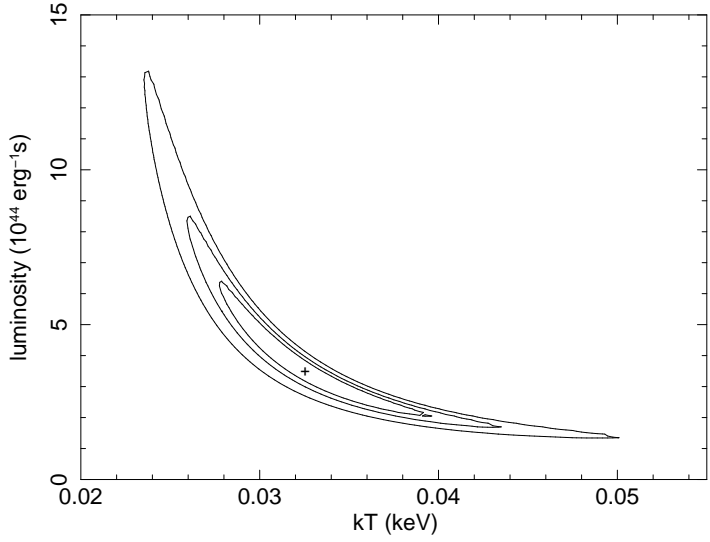}
%\epsscale{0.8}
%\plotone{cont_kt_lum.ps}
\caption{Confidence contours (68\%, 90\%, 99\% for two interesting parameters)
for the fitted effective temperature and
luminosity for the blackbody model component.}
\label{fig:kt-lum}
\end{figure}

Interestingly, the fitted (rest frame) thermal temperatures are surprisingly low,
e.g.\ $kT_{\rm bb}=32.5^{+8.0}_{-6.0}$\,eV for a (redshifted) blackbody.
Such an exceptionally low temperature is
in remarkable contrast to the "canonical" $\sim$0.1--0.2\,keV
found for AGNs \citep{gd04,haba08,bian09},
and is perhaps the lowest with confident measurement known so far\footnote{
Lower $kT$ in a few objects were reported by \citet{urry89} based on
{\it Einstein} IPC+MPC data; however, they were not confirmed in
latter observations, e.g.\ in Mkn\,766  \citep{mol93}.} (to our knowledge).
In fact, the ultra-softness of this component is evident,
as it dominates the emission below  0.4\,keV  only.
The {\em overall} blackbody luminosity is high,
$L_{\rm bb}=3.5^{+3.3}_{-1.5}\times10^{44}$\,\ulum,
whose confidence contours vs.\ $kT_{\rm bb}$
are shown in Figure\,\ref{fig:kt-lum}.
In fact, the blackbody luminosity
is comparable to the "thermal" bolometric luminosity
[$\sim 2.8(\pm 0.6)\times 10^{44}$\,\ulum],
  which is remarkable considering the large uncertainty  in the bolometric correction.
Hence we assume this ultra-soft X-ray emission as blackbody in nature.

The spectrum measured in the RASS has an almost identical spectral shape
as that of the pointed observation.
The best-fit parameters are in excellent agreement with those obtained
from the pointed observation, with $\Gamma=1.47^{+0.77}_{-0.92}$ and
the blackbody temperature $30^{+12}_{-10}$\,eV (90\% confidence level),
though the errors are large given the small source counts.
The overall blackbody luminosity is found to be $3.4\times10^{44}$\,\ulum.
There appears to be no detectable variation in either
the spectral shape or the luminosity between
the RASS and pointed observation.

\section{Emission from an accretion disc?}

\subsection{Size and temperature of X-ray emitting region}

Assuming blackbody radiation, the size of the emitting area can be inferred as
$A=L_{\rm bb}/(\sigma T_{\rm bb}^4)$
($\sigma$ the Stefan-Boltzmann constant).
 For an assumed geometry of a sphere or a face-on disc,
 a radius $\sim5$(or 7)$\times10^{12}$\,cm ($\sim$0.3--0.5\,AU) is inferred,
which is extremely compact for its high luminosity.
 In fact, these radii are merely several to ten times
 the Schwarzschild radius of a BH at the mass estimated for \obj\
 (\rsch$=2GM/c^2\approx 9\times10^{11}$\,cm).
An immediate conclusion is that
the soft X-ray emitting region must be extremely compact,
e.g.\ most likely in the close vicinity of the central BH.

In the standard BH accretion disc model
the local effective temperature at a radius $R$ is
\begin{eqnarray}
T_{\rm eff}\,(R) &=&
\left[\frac{3GM\dot{M}}{8\pi \sigma R^3} \left(1-\sqrt{\frac{R_{\rm in}}{R}}\right)\right]^{1/4}
\\
&=& 6.24\times 10^5 \left(\frac{R}{R_{\rm s}}\right)^{-3/4} \left(1-\sqrt{\frac{R_{\rm in}}{R}}\right)^{1/4}
\left(\frac{M}{10^8 M_{\odot}}\right)^{-1/4}  \dot{m}^{1/4} {\rm (K)}\nonumber
\label{eq:teff}
\end{eqnarray}
 \citep{fkr92,kato08},
where $R_{\rm in}$ is the radius of the inner boundary
(assumed to be the last marginally stable orbit) of the disc, and
$\dot{m}\equiv \dot{M}/\dot{M}_{\rm Edd}$ the scaled accretion rate,
 with the Eddington rate defined as
$\dot{M}_{\rm Edd}\equiv L_{\rm Edd}/(\eta c^2)$,
e.g.\
$1.39\times10^{18}M/M_{\odot}$\,g\,s$^{-1}$ assuming $\eta=0.1$.
For a non-rotating BH ($R_{\rm in}=3R_{\rm s}$),
from outer part inwards $T$  reaches its maximum
\begin{equation}
T_{\rm max}=11.5 \left(\frac{M}{10^8M_{\odot}}\right)^{-1/4} \dot{m}^{1/4} {\rm ~(eV)} ,
\label{eq:tmax}
\end{equation}
at $R=(49/36)\times 3R_{\rm s}$,
and decreases  again inwards.
For \mbh=$3\times 10^6$\,\msun\ and assuming \mdot\ as \redd\ (=0.69),
we find $kT_{\rm max}$ = 25.3\,eV, that is close to
the fitted \tbb$=32.5^{+8.0}_{-6.0}$\,eV.
To our knowledge, this is by far the most convincing case where the
 soft X-ray excess temperature is well compatible
with the predicted maximum disc temperature.

\subsection{Accretion disc model fit}
\subsubsection{Standard accretion disc model and parameters}

We fit the spectrum with
the commonly adopted multi-colour disc
(MCD, {\it diskbb} in {\it XSPEC}) model
\citep[][]{mit84,mak86}.
This model has been widely used in fitting the X-ray spectra of BHXBs,
and found to provide reasonably good descriptions
to the observed spectra at the high/soft state
\citep[e.g.][]{dot97,kub98}.
With an approximation to the boundary condition,
the MCD model has a monotonic temperature profile,
$T_{\rm eff}\,(R)= T_{\rm in}(R/R_{\rm in})^{-3/4}$,
where  $T_{\rm in}$ is the effective temperature at the disc inner radius $R_{\rm in}$.
This model should be sufficient for our purpose here,
considering the moderate spectral S/N and resolution and the
 limited bandpass of the PSPC data.

The emergent MCD spectrum is determined by two parameters:
the (apparent) radius of the inner boundary of the disc
$R'_{\rm in}$ and the maximum (colour) temperature \tinc,
which are related to the disc luminosity via
\begin{equation}
L = 4\pi R'^2_{\rm in} \sigma T'^4_{\rm in} .
%L = 4\pi R'^2_{\rm in} \sigma T'^4_{\rm in} = 4\pi D^2 f /\cos\theta .
\label{eq:discbblum}
\end{equation}
\tinc\ and \rinci\ can be derived from spectral fitting,
where $\theta$ is the disc inclination angle.
Since the MCD model does not incorporate the redshift effect,
which is small for \obj\ at $z$=0.116 though,
\tinc\ and \rinc\ in the object's rest frame are obtained
by multiplying the fitted values with ($1+z$) and
$1/(1+z)$ (can be shown by using Eq.\,\ref{eq:discbblum}), respectively.
The {\em true} inner boundary radius \rin\
is related to the apparent radius $R'_{\rm in}$ via
$R_{\rm in}= \xi R'_{\rm in}$, where
$\xi=\sqrt{3/7}(6/7)^3 \approx0.41$ is
introduced to account for the inner boundary
condition that is neglected in the MCD model \citep{kub98,kato08}.

The disc inclination $\theta$ is expected to be a small angle
given the blazar-like property of \obj,
i.e.\ generally $\la 10\arcdeg$ assuming disc--jet perpendicularity.
In fact, for a wider range of $\theta$, say $\theta < 30\arcdeg$
(as expected for type\,1 AGNs),
the dependence of $\sqrt{\cos\theta}$  on $\theta$ is weak.
We thus expect
 $\sqrt{\cos\theta}\approx 1$, and hence \rini$\approx$\rin.

To relate the observed spectrum to the true disc parameters
 corrections should be made to account for the spectral hardening effect,
which gives rise to a shifted Wien spectrum to a higher colour temperature
due to incoherent Compton scattering \citep{ce87,wp88,ross92,st95a}.
The true effective \tmax\ is then
\begin{equation}
T_{\rm max} = \frac{1}{\kappa} T'_{\rm in},
\label{eq:t_corr}
\end{equation}
where $\kappa$ is the spectral hardening factor.
Consequently, the true inner boundary radius
\begin{equation}
R_{\rm in}=  \kappa^2  \xi R'_{\rm in}
\label{eq:r_corr}
\end{equation}
 \citep{kato08,kub98}.
The value of $k$ is somewhat uncertain, and
$\kappa \simeq 1.7\pm0.2$ was suggested \citep{st95b}
based mostly on consideration of BHXBs
and $1<\kappa\la 2.5$ for SMBHs with wide ranges
of \mbh\ and \mdot\ \citep{ross92,st95a}.
Here we adopt $\kappa \simeq 1.7$.

\subsubsection{Results of accretion disc model fit}

\begin{figure}
\epsscale{0.8} \plottwo{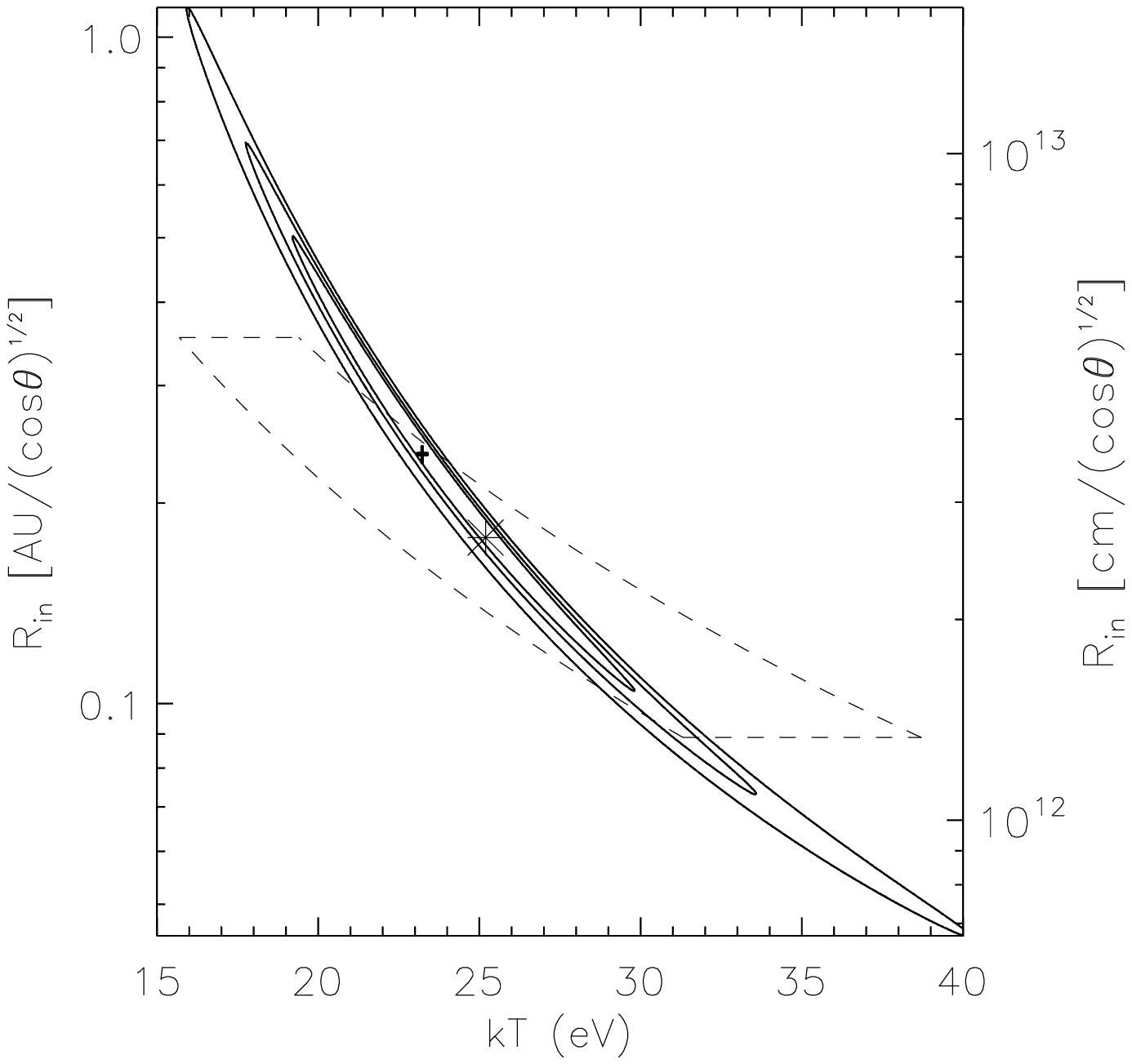}{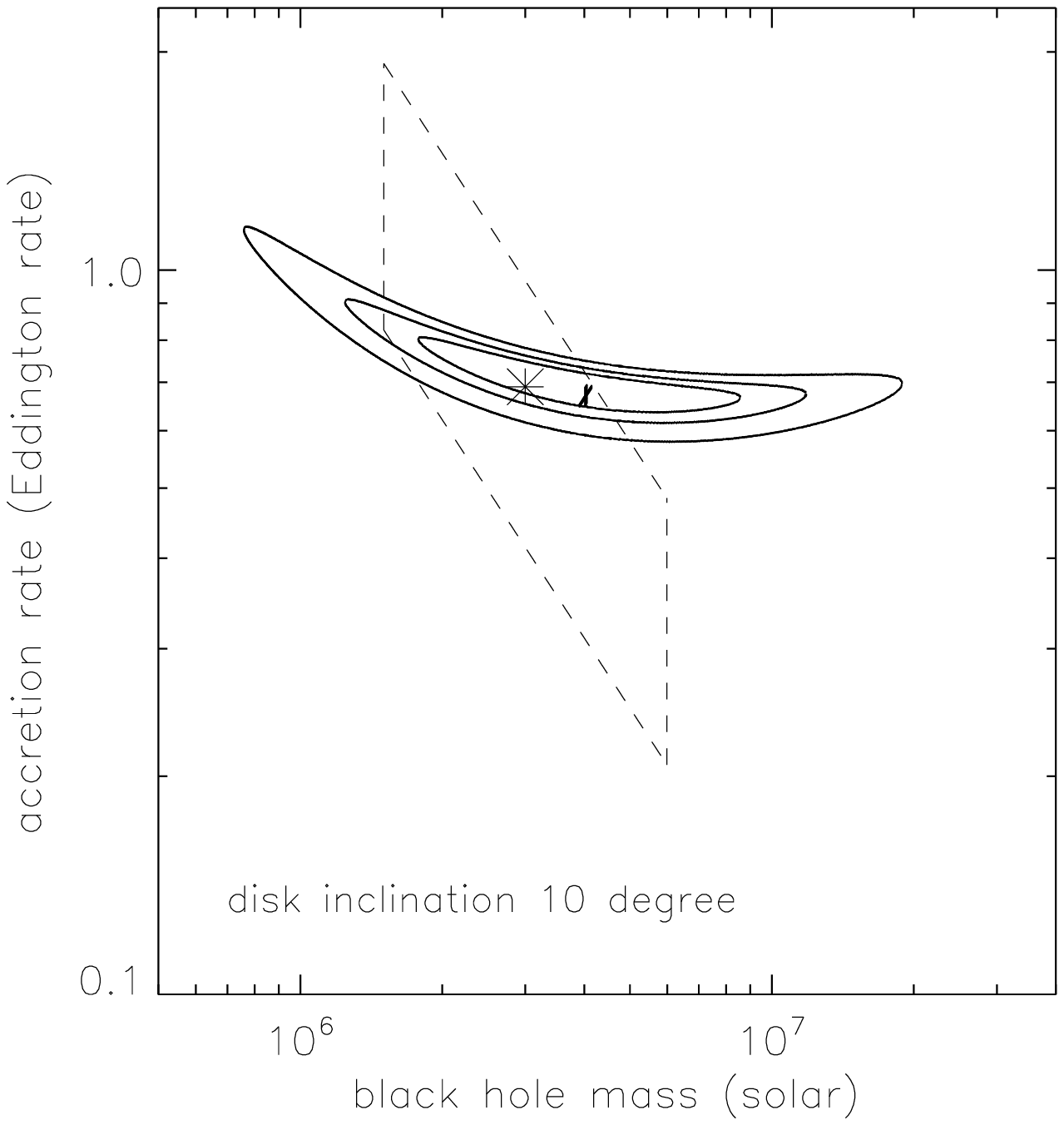}
\caption{Confidence contours (solid; 68\%, 90\%, 99\%) for the
maximum effective temperature vs.\ the inner radius of the disc
(left panel) and the BH mass versus scaled accretion rate (right
panel; assuming a disc inclination $\theta=10\arcdeg$), derived from
fitting the spectrum with the MCD plus a power-law  model. The
estimated values (asterisk) derived from the optical emission line
spectrum
 and their uncertainty ranges  are indicated
(dashed; $\pm$0.3dex in BH mass and
$\pm2(1\sigma)$ in the bolometric correction factor are used).
}
\label{fig:kt-rin}
\end{figure}

The spectrum is well fitted with the MCD plus a power-law model,
with the fitted \nh\ consistent with \gnh.
Fixing \nh=\gnh\ yields the same excellent fit as using blackbody,
giving
 \ktinc=$39.6^{+12.3}_{-7.5}$\,eV and
\rinc=$3.0^{+2.9}_{-1.4}\times10^{12}/\sqrt{\cos\theta}$\,cm
in the object's rest frame  (Table\,1).
Applying the corrections in Eqs.\,\ref{eq:t_corr} and \ref{eq:r_corr}
with $\kappa\simeq 1.7$,
we find a true maximum effective temperature
\ktmax=$23.3^{+7.2}_{-4.4}$\,eV
and an inner disc radius
\rin=$3.6^{+3.4}_{-1.7}\times10^{12}/\sqrt{\cos\theta}$\,cm.
The confidence contours of  \ktmax\ and \rin\
are shown in Figure\,\ref{fig:kt-rin}.

Assuming a non-rotating BH and \rin$=3$\rsch,
we can derive its mass from X-rays, \mbhx,
and in turn \mdot\ using \tmax\ and  Eq.\,\ref{eq:tmax}.
We find the best estimates
\mbhx$=4.1\times10^6/\sqrt{\cos\theta}$\,\msun\ and
\mdot$=0.68/\sqrt{\cos\theta}$.
The contours of confidence intervals for \mbh\ and \mdot,
as derived from those of \rin\ and \tmax\ assuming $\theta=10\arcdeg$,
 are shown in Figure\,\ref{fig:kt-rin} (right panel).
It is interesting to note that the X-ray spectrum provides relatively
tight constraint on accretion rate \mdot.
This is because the uncertainties of the derived \rin\ (\mbhx) and \tmax,
albeit large, are highly inversely coupled (left panel)
and hence largely canceled mutually.

These derived parameter values can be compared with those
estimated from the optical spectrum,
 \mbh\ and \redd\ (assumed as \mdot),
 and in turn \rin\  and \tmax\
 (using Eq.\,\ref{eq:tmax} and assuming \rin=3\rsch).
Their best estimates and
uncertainty ranges are overplotted in Figure\,\ref{fig:kt-rin},
as $\pm$0.3\,dex in \mbh\ \citep{vp06} and $\pm2(1\,\sigma )$
in the bolometric correction \citep{rich06}.
Remarkably, the results derived from
the X-ray and optical observations agree strikingly well,
albeit the relatively large uncertainties in both the measurements.
This conclusion is basically not affected by the exact value of $\kappa$,
provided $1\la \kappa \la 2.5$.

We note that the MCD model is only approximately correct since
no general relativistic (GR) effects are taken into account.
We make further corrections to the derived disc parameters
due to the GR effects following the approach of \citet{zhang97}.
Eqs.\ref{eq:t_corr} and \ref{eq:r_corr} are then
$T_{\rm max} = f^{-1} (\kappa^{-1} T'_{\rm in})$
and
$R_{\rm in}= g^{-0.5}f^2 (\kappa^2  \xi R'_{\rm in})$,
where $f$ and $g$ are temperature and flux correction factors, respectively.
For a non-rotating BH and $\theta\approx 0\arcdeg$,
$f\approx0.85$ and $g\approx0.80$ \citep[][their Table\,1]{zhang97}.
Therefore the above \ktmax\ and \rin\ (\mbhx) values
should be further multiplied by
a factor of 1.17 and 0.81, respectively,
i.e.\ \ktmax=$27$\,eV,
\rin$\sim2.9\times10^{12}$\,cm,
\mbhx$\sim 3.3\times10^6$\,\msun, and \mdot$\sim1.0$.
These parameter values are still marginally consistent with
those derived from the optical spectrum (Figure\,\ref{fig:kt-rin}).

Therefore, we conclude that we might be seeing the direct X-ray emission
from an accretion disc around the SMBH in \obj,
which has been long sought for AGNs.
We note that, however, for a more realistic modeling,
a relativistic disc model taking the GR effects fully into account
should be used (such as the kerrbb model of Li et al.\ 2005).
Besides, we did not consider the BH spin here, that is also plausible;
e.g.\ for a spinning BH and a prograde disc,
\rin\ would extend further inwards within 3\,\rsch, i.e.\ \rin$<$3\,\rsch,
and \tmax\  increases, resulting in an increased \mbhx\ and a reduced \mdot.
Furthermore, since the above found \mdot$>$0.3,
the effects of a slim disc
(Abramowicz et al.\ 1988, Sadowski 2009, Li et al.\ 2010,
see Chapter\,10 of Kato et al.\ 2008 for a review)
may start to set in.
We defer all these comprehensive treatments to future work,
in a hope that it may shed light on the spin of the BH.

\section{Discussion}
\subsection{Compatibility with lower-frequency data}

\begin{figure}
\epsscale{0.7}
\plotone{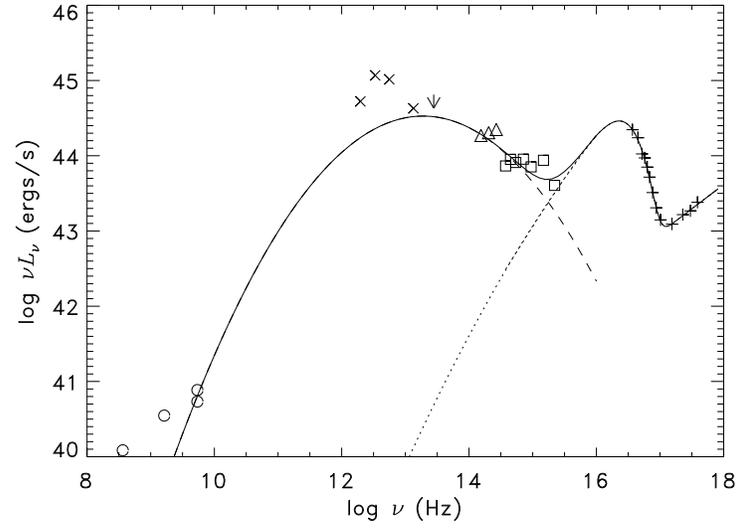}
\caption{Broad-band SED of \obj:
the unfolded PSPC data (crosses) and
the photometric measurements in the UV and  optical ({\it Galex} and SDSS, squares)
near-IR (2MASS, triangles), far-infrared
(\iras\ and \iso, crosses; arrows indicate upper limits) and radio (circles) bands.
The PSF magnitudes are used whenever available.
The models are the best-fit MCD+power-law model (dotted), synchrotron emission (dashed)
and the sum (solid).
}
\label{fig:sed}
\end{figure}

Having established the accretion disc emission model from
the soft X-ray spectrum,
it would be interesting to check whether this model
is compatible with the observed broad-band spectral energy distribution (SED)
of \obj.
The unfolded PSPC spectrum and the fitted MCD plus power-law model (dotted line)
are  shown in Figure\,\ref{fig:sed}, along with
the broad-band SED whose low-energy data are collected
from the NASA/IPAC Extragalactic Database (NED).
Interestingly, while the UV-to-infrared spectrum steepens towards lower frequencies,
 the {\it Galex} far-UV measurement (squares) agrees roughly with the MCD
model prediction.
The lack of an obvious big-blue-bump in  optical-UV supports
that it is likely shifted to higher frequencies.
As for the optical--infrared bump,
we interpret it as the high-energy tail of beamed synchrotron emission
from the relativistic jets,
in light of the  blazar-like property of \obj.
We model this component with a parabolic function,
 a commonly used approximation to the synchrotron emission of blazars.
The sum of this model and the above best-fit MCD model
 is fitted to the radio 5\,GHz, near-IR and optical/UV data,
excluding the far-IR measurements from \iras\ and \iso\ (crosses;
since they are most likely seriously contaminated, or even dominated
by thermal emission of dust\footnote{
The observed $f(60\mu)/f(100\mu)$=0.53 is
"infrared warm", typical of starburst in the local universe \citep{soi87}.
} from the companion starburst galaxy 4\arcsec away
given the relatively poor spatial resolution of \iras\ and \iso).
As can be seen, such a synchrotron component (dashed line)
can generally account for the SED at lower energies.
Therefore, the above accretion disc model,
together with the other two emission components (synchrotron and inverse-Compton),
is compatible with the observed broad-band SED of \obj.
Moreover, the lack of detectable variation of the ultra-soft X-ray emission
between the RASS and pointed observations also supports the accretion disc model,
since the thermal emission from an accretion disc is generally stable
over such a relatively short timescale.

We note that the starburst companion must fall within the source extraction region
on the PSPC detector plane due to its low spatial resolution.
Starburst also produces soft X-rays, which are related to the far-IR
luminosity as $L_{\rm FIR}^{\rm SB}/L_{\rm 0.3-2keV}=10^{3-4}$.
For the companion starburst galaxy,
an upper limit on $L_{\rm FIR}^{\rm SB}$ can be set
by the \iras\ measurements
using $f_{100\mu}=1.12$\,Jy and $f_{60\mu}=0.595$\,Jy \citep{sm96},
yielding
$L_{\rm FIR}^{\rm SB}<L_{\rm FIR}=3.0\times10^{11}\,L_{\odot}$
($ 1.15\times10^{45}$\,\ulum).
This corresponds to an upper limit on the soft X-ray luminosity
arising from the starburst
$L_{\rm 0.3-2keV}^{\rm SB}<1\times 10^{41-42}$\,\ulum\ only.
Such a contribution, if any, would be
overwhelmed by the X-ray emission from the AGN given
the observed luminosity $L_{\rm 0.3-2keV}\sim 10^{44}$\,\ulum.
Moreover, the thermal temperatures of X-ray gases of starburst
fall within a strict range of 0.3--0.7\,keV, far higher than the
fitted 50\,eV for a thin plasma model (Table\,1).
We thus conclude that the \rosat\ spectrum of \obj\ is essentially
not affected by any potential contamination from the starburst companion.

\subsection{Why is \obj\ different from other AGNs?}

An interesting question arises as to why \obj\ is so special and unique,
in terms of previous unsuccessful attempts to ascribe the soft X-ray excess
(with the canonical temperatures of $\sim$0.1--0.2\,keV)
to the direct thermal emission of the accretion discs in the vast majority of AGNs.
This question will not be answered until the origin of the puzzling
soft X-ray excess in many other AGNs is understood.
Nonetheless, we discuss some possibilities here, albeit somewhat speculative.
In any case, one important feature of \obj\ is its dominance of the disc thermal emission,
which is a close analogue to the typical high/soft state in BHXBs
(relatively weak corona; see above).
This is different from many other AGNs in which the soft excess was investigated,
where the spectral state is characterised by strong corona emission.

One of the viable explanations of the soft excess
is Comptonisation of soft photons (e.g. from discs)
off a region with electron temperature $kT_{\rm e}\sim 0.1-0.2$\,keV
and a large optical depth $\tau\sim20$ \citep[e.g.][]{ce87,gd04,done07b}.
In this scenario, \obj\ may differ from the other AGNs in its environment
surrounding the central engine, i.e. lacking this Comptonisation region
and the disc being directly seen in the line-of-sight,
which may not be the case for the other AGNs.
For instance, the difference may be due to the presence of jets
(and the face-on inclination of the disc) in \obj.
It may be possible that in radio-quiet AGNs
 the Comptonisation region is formed by aborted jets \citep{hp97,ghi04}
whose kinetic energy is converted to
internal energy to heat electrons to the required temperature.
In the case of \obj, the jets are successfully launched, however,
clearing the way to the central disc along the line-of-sight.
It is interesting to note that in 3C\,273, a well-known AGN with relativistic jets,
a relatively low blackbody temperature of the soft excess was also reported
($\sim 60$\,eV; Grandi \& Palumbo 2004; although in that work the accretion disc
origin was not tested in terms of matching the X-ray data with
the predicted temperature and luminosity
of the disc, as we did here).
Whether or not this can be generalised to other radio-loud AGNs needs further
investigations, though it should be reminded that
the jets of radio-loud NLS1s might be different
from those of classical radio quasars, e.g.\ with reduced physical size and power
 \citep{yuan08}.

An alternative model is disk reflection in which
the observed soft X-ray excess can be explained as
relativistically blurred line emission of the reflection component
from a highly ionized inner disc, that somehow dominates the primary emission
 \citep[][]{ross93,ball01,mini04,crum06}.
The postulation of this model is the presence of the primary X-ray emission,
presumably from a disc corona of high-temperature and low-optical depth.
It may be possible  that in \obj\ the disc corona is significantly suppressed somehow,
for instance, due to the formation of the jets (the energy goes into the jets
instead of to the corona) or to the relatively high mass accretion rate.
This would lead to a week reflection component in \obj\ and thus has no significant
soft excess in the usual form, while the disc thermal emission stands out.
In this line,
as mentioned above, the Seyfert-like X-ray component in the \rosat\ spectrum
of \obj\ is too weak to be detected.

Finally, for AGNs with high black hole masses ($M> 10^7$\,\msun)
and relatively low accretion rates,
even if the other conditions (jets, corona, etc.)
resemble those of \obj, the thermal disc emission may have too low temperatures
to be detectable above the low-energy cutoff of commonly used
X-ray detectors ($\sim 0.1$\,keV).
Future observations of more AGNs similar to \obj\ may shed light on under what
conditions the direct thermal emission of accretion discs
can be detected in X-ray.

\section{Concluding remarks}

We have shown that the ultra-soft X-ray excess discovered in \obj\
 can be well and self-consistently described by thermal emission
 from an optically thick accretion disc around a SMBH.
The derived parameters
(\mbhx, \mdot, and \tmax) agree with the
independent estimates based on the optical spectrometric data.
We thus consider this emission as a signature of X-rays from
 an accretion disc around a SMBH,
 presenting possible evidence for BH accretion discs in AGNs.
As such, \obj\ is an AGN analog of BHXBs at the high/soft state
in a sense that the accretion disc emission dominates the bolometric luminosity.
More importantly,
one may infer the  spin of the BH,
as succeeded in BHXBs \citep[e.g.][]{zhang97,mcc06}.
This may be achieved with future improved observations,
which, if successful, would provide a testbed for studying
 the link of BH spin and the formation of relativistic jets in AGNs.
Finally, it is intriguing why \obj\ is so unique among the
many AGNs observed in the soft X-rays, which may be
addressed with  future investigations.

\acknowledgments
WY thanks S.\ Mineshige, A.C.\ Fabian, and R.\ Taam
for reading the manuscript and providing helpful comments,
and F.\ Meyer and E.\ Meyer-Hofmeister for discussion,
as well as L.M. Dou and X.B. Dong for help with some of the data preparation.
This work is supported by NSFC grants 10533050, 10773028,
the National Basic Research Program of (973 Program) 2009CB824800.
This research has made use of the ROSAT all-sky survey data and
pointed observation data which have   been processed at MPE.
This research has made use of the SDSS data,
%This research has made use of
and the NASA/IPAC Extragalactic Database (NED)
which is operated by the Jet Propulsion Laboratory, California Institute of Technology,
under contract with the National Aeronautics and Space Administration.

\begin{deluxetable}{lcccc}
\tablenum{1} \tablewidth{0pt} \topmargin 0.0cm \evensidemargin = 0mm
\oddsidemargin = 0mm
%%\rotate
%\tabletypesize{\tiny}
\tablecaption{
Results of spectral fits
}
\tablehead{ \colhead{wabs * model\tablenotemark{~a}} &
            \colhead{\nh ($10^{20}$\,\unh)} &
            \colhead{$\Gamma$} &
            \colhead{$kT$(eV) / $\Gamma_{\rm s}$\tablenotemark{b}} &
            \colhead{$\chi^2$/d.o.f.}
}
\startdata
powl & $0.37^{+0.30}_{-0.25}$ & $2.22^{+0.22}_{-0.20}$ & & 34.1/29 \\
powl & 1.79(fixed)\tablenotemark{c} & 3.06 &  & 67.4/30 \\
powl (0.4--2.4\,keV) & 1.79(fixed) & $1.63\pm0.40$ & & 4.4/9 \\
powl + powl    & 1.79(fixed) & $1.04^{+0.57}_{-0.52}$ & $4.83^{+1.53}_{-0.56}$ & 17.2/28\\
powl + raymond & 1.79(fixed) & $1.30^{+0.42}_{-0.43}$ & $50^{+15}_{-11}$ & 16.4/28 \\
powl + zbremss & 1.79(fixed) & $1.34^{+0.42}_{-0.40}$ & $59^{+26}_{-15}$ & 16.7/28 \\
powl + zbbdy  & $1.1^{+4.2}_{-0.9}$ & $1.30^{+0.42}_{-0.64}$ & $42^{+41}_{-21}$& 16.6/27\\
powl + zbbdy  & 1.79(fixed)& $1.37\pm0.49$ & $32.5^{+8.0}_{-6.0}$ & 16.9/28\\
powl + zbbdy (RASS) & 1.79(fixed)& $1.47^{+0.77}_{-0.92}$ & $30^{+12}_{-10}$ & 10.7/9 \\
powl + diskbb & $1.2^{+2.5}_{-0.7}$ & $1.29^{+0.40}_{-0.66}$ & $48^{+53}_{-27}$ & 16.6/27 \\
powl + diskbb & 1.79(fixed) & $1.36\pm0.41$ & $39.6^{+12.3}_{-7.5}$ & 16.8/28
\enddata
\tablenotetext{a}{Source model (as in XSPEC)
modified by neutral absorption (wabs) with column density \nh:
powl -- power-law with photon index $\Gamma$;
raymond -- Reymond-Smith thin plasma emission (abundance fixed at solar);
zbremss -- redshifted thermal bremstrahlung;
zbbdy -- redshifted blackbody with effective temperature $kT$;
diskbb -- multi-colour disc model with the (rest frame) maximum temperature $kT$ (see text)}
\tablenotetext{b}{$\Gamma_{\rm s}$: photon index of a second softer power-law}
\tablenotetext{c}{the Galactic \nh\ value, fixed in fitting}
\end{deluxetable}

\end{document}